\documentclass[aps,prx,reprint,twocolumn,superscriptaddress,nofootinbib,longbibliography,floatfix]{revtex4-2}

\usepackage{graphicx}
\usepackage{amsmath}
\usepackage{braket}
\usepackage{bm}
\usepackage{xcolor}
\usepackage[colorlinks=true,linkcolor=blue!60!black,citecolor=blue!60!black,urlcolor=blue!60!black]{hyperref}

\graphicspath{{Figures/}{Supplementary_Figures/}}

\begin{document}

\title{Quantum jumps of sound}

\author{Takuma~Makihara}
\thanks{These authors contributed equally to this work.}
\affiliation{Department of Applied Physics and Ginzton Laboratory, Stanford University, Stanford, California 94305, USA}

\author{Erik~Szakiel}
\thanks{These authors contributed equally to this work.}
\affiliation{Department of Applied Physics and Ginzton Laboratory, Stanford University, Stanford, California 94305, USA}

\author{Matthew~P.~Maksymowych}
\affiliation{Department of Applied Physics and Ginzton Laboratory, Stanford University, Stanford, California 94305, USA}

\author{Oliver~A.~Hitchcock}
\affiliation{Department of Physics, Stanford University, Stanford, California 94305, USA}

\author{Kaveh~Pezeshki}
\affiliation{Department of Applied Physics and Ginzton Laboratory, Stanford University, Stanford, California 94305, USA}

\author{Rachel~G.~Gruenke-Freudenstein}
\affiliation{Department of Applied Physics and Ginzton Laboratory, Stanford University, Stanford, California 94305, USA}

\author{Mihir~Pendharkar}
\affiliation{Department of Applied Physics and Ginzton Laboratory, Stanford University, Stanford, California 94305, USA}
\affiliation{Department of Physics, Stanford University, Stanford, California 94305, USA}

\author{Shannon~P.~Harvey}
\affiliation{SLAC National Accelerator Laboratory, Menlo Park, California 94025, USA}

\author{David~I.~Schuster}
\affiliation{Department of Applied Physics and Ginzton Laboratory, Stanford University, Stanford, California 94305, USA}
\affiliation{Department of Physics, Stanford University, Stanford, California 94305, USA}

\author{Amir~H.~Safavi-Naeini}
\email{safavi@stanford.edu}
\affiliation{Department of Applied Physics and Ginzton Laboratory, Stanford University, Stanford, California 94305, USA}

\date{September 17, 2026}

\begin{abstract}
Quantum mechanics predicts that a vibrating object's energy comes in discrete packets, yet no measurement of its position reveals this discreteness. Resolving individual energy levels requires a qualitatively different measurement, one coupling to the resonator's energy rather than its displacement. We use a superconducting qubit dispersively coupled to a nanomechanical resonator to perform repeated quantum nondemolition measurements of the phonon number. An aligned transfer-print technique integrating the qubit and resonator yields a mechanical lifetime of $T_1 = 2.1$~milliseconds and a dispersive shift of $2\chi/2\pi = 328$~kilohertz per phonon. We heralded single-phonon states with $85\%$ fidelity and observed quantum jumps between the resonator's first excited state and ground state. These discontinuous transitions are a striking manifestation of quantum mechanics in a massive, vibrating object.
\end{abstract}

\maketitle

\begin{figure}
\centering
\includegraphics[width=\columnwidth]{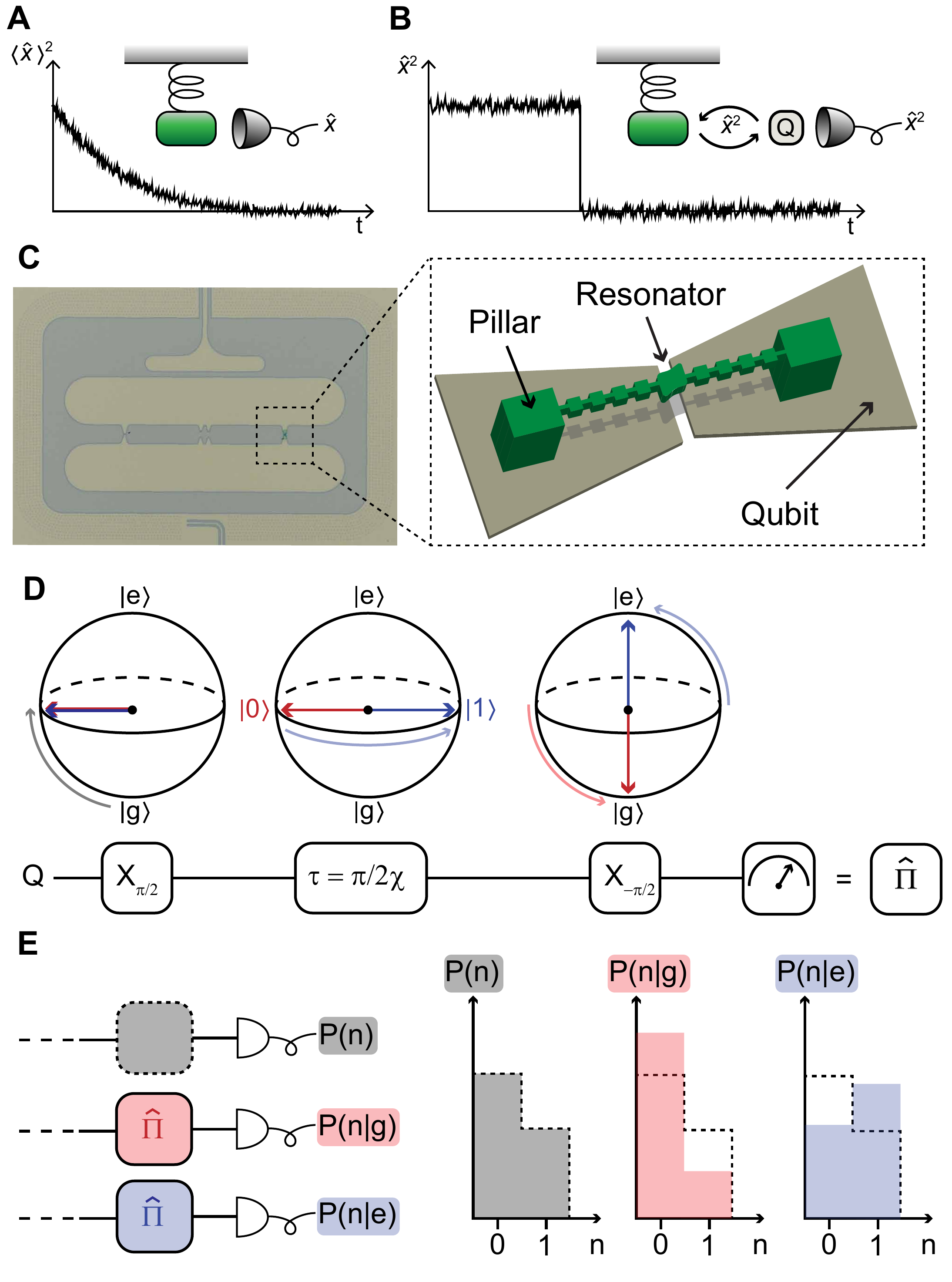}
\caption{\label{fig:device} \textbf{Device architecture.} \textbf{(A)} The quantization of sound is hidden from an observer measuring the position $\hat{x}(t)$ of a mechanical resonator. If an observer measures $\braket{\hat{x}(t)}$ and calculates the energy of the resonator as $\braket{\hat{x}(t)}^2$, a continuous exponential decay results. \textbf{(B)} If an observer can directly measure $\hat{x}^2(t)$ of a mechanical resonator, then quantum jumps in the mechanical energy can be observed. By using a qubit and mechanical resonator coupled in the dispersive regime, one can directly measure $\hat{x}^2(t)$. \textbf{(C)} Optical micrograph of the device measured in this work, consisting of a transmon qubit coupled to a LN mechanical resonator. (Inset) Illustration of the qubit-mechanics integration. A one-dimensional phononic crystal cavity of LN (shown in green) is suspended by pillars such that it floats 75 nm above the qubit islands. \textbf{(D)} Qubit pulse sequence used to measure the parity of the LN mechanical resonator. By exciting the qubit Bloch vector onto the equator and letting the qubit and mechanics dispersively interact for $\tau=\pi/2\chi$, one can measure the parity of the mechanical resonator. We summarize a Ramsey pulse sequence and qubit state measurement by the operator $\hat{\Pi}$. \textbf{(E)} After preparing the mechanical resonator in an initial coherent state, each application of $\hat{\Pi}$ and constituent qubit measurement outcome modifies the underlying phonon number distribution $P(n)$. We measure $P(n)$ through qubit spectroscopy. In our pulse sequence, measuring the qubit in $\ket{g}$ ($\ket{e}$) increases the distribution weight in even (odd) phonon numbers.}
\end{figure}

Measurements that couple to a vibrating object's displacement are fundamentally unable to resolve the discrete energy levels predicted by quantum mechanics. For a resonator starting in an initially excited state, the expected value $\braket{\hat{x}(t)}^2$ reveals only a smooth exponential decay (Fig.~\ref{fig:device}A), identical to that of a classical oscillator. To observe the quantized nature of mechanical energy, one must instead measure the energy directly, avoiding the back action that displacement measurements impart on the oscillator's momentum~\cite{braginsky1980quantum}. Such a quantum nondemolition (QND) measurement of $\hat{x}^2$ (Fig.~\ref{fig:device}B) requires coupling the oscillator to an intrinsically nonlinear system.

Discontinuous jumps between quantized energy levels were first proposed by Bohr~\cite{bohr1913constitution} and first directly observed in single trapped ions~\cite{bergquist1986observation,nagourney1986shelved}. For harmonic oscillators, quantum jumps between photon number states were observed in electromagnetic cavities, first in a Fabry-Perot cavity probed by Rydberg atoms~\cite{gleyzes2007quantum,guerlin2007progressive} and later in a three-dimensional microwave cavity coupled to a superconducting qubit~\cite{sun2014tracking}. In each case, the critical ingredient was a nonlinear detector that could measure the photon number without disturbing it. Extending these observations to mechanical systems is a compelling goal: Mechanical resonators are massive objects whose vibrations we intuitively expect to be continuous, making them a remarkable setting in which to observe quantum discreteness. They are already central to precision measurement. For example, kilogram-scale pendulums are used for gravitational wave detection~\cite{abbott2016observation}. By contrast, integrated electromechanical resonators, like those studied here, are exceptionally sensitive to small masses and forces. In particular, their small effective mass makes them particularly well-suited for integrated mass spectrometry and probing single atomic defects~\cite{naik2009towards,hanay2012single,yuksel2026intrinsic,kuruma2025controlling}. Furthermore, their compact size, long lifetimes, and low cross-talk make them promising for quantum information processing~\cite{pechal2018superconducting,chamberland2022building}. Their ability to couple to diverse physical degrees of freedom also makes them natural candidates for quantum transduction. However, observing quantum jumps in a mechanical resonator demands a long mechanical lifetime, a strongly coupled nonlinear detector, and fast, high-fidelity measurements. This combination has not previously been realized.

\section*{Integrated device architecture}
Coupling piezoelectric mechanical resonators to superconducting qubits~\cite{o2010quantum,chu2017quantum} has enabled measurements of quantized mechanical energy levels~\cite{arrangoiz2019resolving,von2022parity,lee2023strong}, preparation of non-Gaussian mechanical states~\cite{wollack2022quantum,chu2018creation,bozkurt2025mechanical,bild2023schrodinger}, and quantum information transfer via phonons~\cite{bienfait2019phonon,bienfait2020quantum,qiao2023splitting}. However, tracking phonon number in real time requires simultaneously achieving a long mechanical lifetime, strong dispersive coupling, and good qubit coherence, which has been limited by the difficulty of integrating piezoelectric materials with superconducting circuits without degrading either system~\cite{jain2023acoustic}.

Our nanomechanical resonator is a one-dimensional phononic crystal cavity fabricated from thin-film lithium niobate (LN)~\cite{arrangoiz2019resolving,wollack2022quantum,lee2023strong}. A micron-scale central rectangular block is suspended from both ends by periodically patterned beams that are engineered to have a full phononic bandgap, as illustrated in the inset of Fig.~\ref{fig:device}C. These periodic structures act as acoustic mirrors that localize mechanical modes of the central block and protect them from phononic radiation into the bulk. The key design challenge is to maximize the overlap between the qubit's electric field and the mechanical mode while minimizing interaction with any additional piezoelectric volume that may lead to loss~\cite{jain2023acoustic}. We fabricate the mechanics and qubit separately to incorporate state-of-the-art techniques that are incompatible across material systems and then integrate them by using an aligned transfer print process (Appendix~\ref{app:fabrication}). We design the qubit with a narrow 500-nm gap between its islands, which concentrates the electric field for efficient coupling. Using a submicron die bonder, we align and stamp the fully fabricated phononic crystal onto this coupling region. Raised pillars of LN on either end of the phononic crystal act as spacers, resulting in the resonator being suspended just 75 nm above the qubit gap (Fig.~\ref{fig:device}C).

The integrated device achieves a phonon lifetime of $T_1 = 2.1$ ms, exceptionally long for any piezoelectric resonator at gigahertz frequencies, and a dispersive shift of $2\chi/2\pi = 328$ kHz, yielding a measurement-to-lifetime ratio that allows roughly 170 parity checks within a single phonon lifetime. The qubit maintains good coherence ($T_2^* = 3.2~\mu$s, $T_{1q} = 31~\mu$s) despite direct contact with the piezoelectric. Full device parameters are given in Table~\ref{table:sup_device_params}.

\begin{figure*}
\centering
\includegraphics[width=\textwidth]{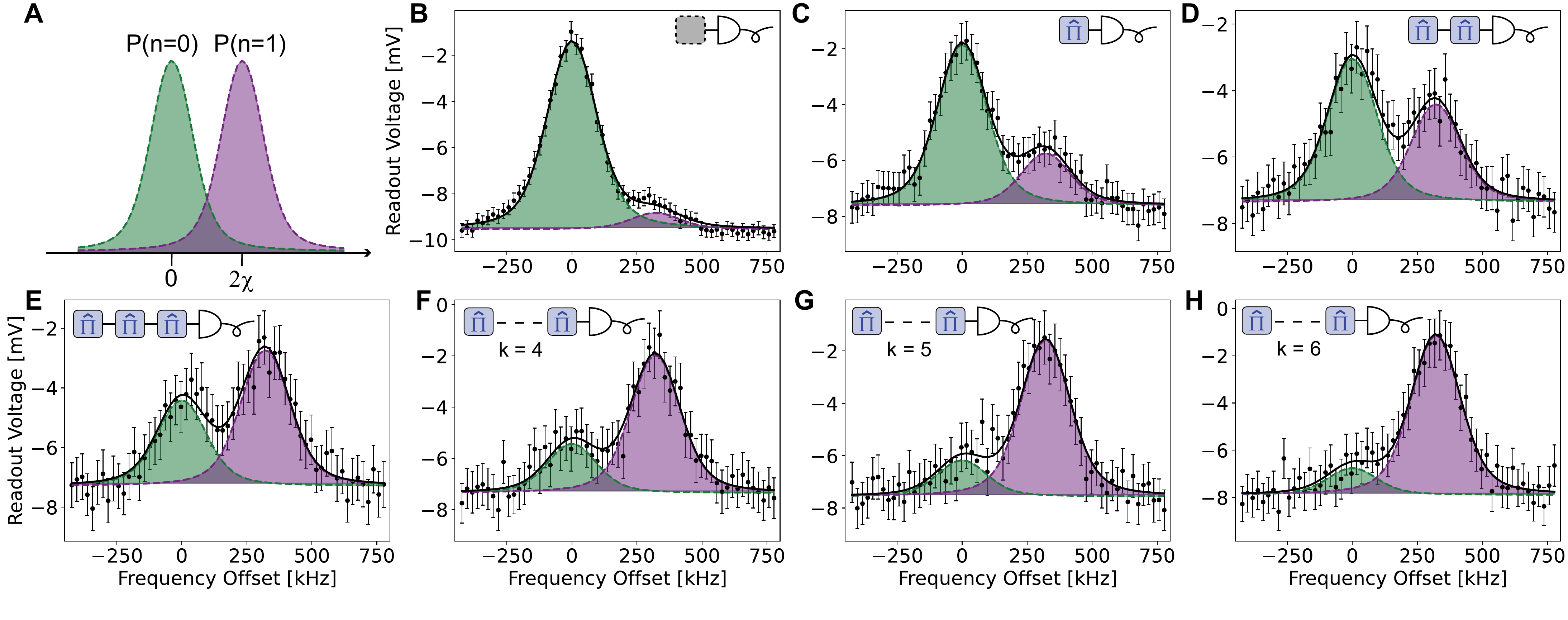}
\caption{\label{fig:spectra} \textbf{Postselected qubit spectra.} \textbf{(A)} Illustration of the qubit excitation spectrum in the strong dispersive regime. The area under the peak centered at zero frequency offset ($2\chi$ frequency offset) is proportional to the probability of measuring $n=0$ ($n=1$) phonons in the mechanical resonator. \textbf{(B)} Qubit spectrum measured with no parity checks, representing the initial coherent state of the mechanics. \textbf{(C to H)} Qubit spectrum postselected on $k=1, 2, 3, 4, 5, 6$ consecutive QND parity measurements, where every measurement outcome is $\ket{e}$. As we observe more consecutive measurements of the qubit in the excited state, we become more certain that our mechanical resonator is in the $n=1$ state, corresponding to a growing weight in the peak with frequency offset $2\chi/2\pi=328$ kHz.}
\end{figure*}

We operate the system in the dispersive regime, where the qubit and mechanics are far detuned and the effective Hamiltonian is~\cite{koch2007charge}
\begin{equation}
\hat{H}_\text{eff} = \omega_\text{m} \hat{b}^\dagger \hat{b} + \frac{1}{2}(\omega_\text{ge} + 2\chi \hat{b}^\dagger \hat{b}) \hat{\sigma}_z \: 
\end{equation}
with $\hat{b}$ the phonon lowering operator and $\hat{\sigma}_z$ the qubit Pauli operator. The qubit transition frequency shifts by $2\chi$ for each phonon in the resonator. Crucially, the phonon number $\hat{n} = \hat{b}^\dagger\hat{b}$ commutes with $\hat{H}_\text{eff}$, so measurements of the qubit frequency are QND with respect to the phonon number~\cite{braginsky1980quantum,gleyzes2007quantum,sun2014tracking}.

To implement this QND measurement, we use a Ramsey pulse sequence that maps the phonon parity onto the qubit state (Fig.~\ref{fig:device}D). A $X_{\pi/2}$ pulse prepares the qubit on the equator of the Bloch sphere, where it precesses about the $Z$ axis at a rate $2\chi n$. After a time $\tau = \pi/2\chi = 1.524~\mu$s, even phonon numbers return the Bloch vector to its starting position while odd phonon numbers accumulate a phase of $\pi$. A final $X_{-\pi/2}$ pulse and measurement then map even (odd) phonon numbers to $\ket{g}$ ($\ket{e}$). We denote this parity check by the operator $\hat{\Pi}$. Because we prepare initial mechanical states with $\bar{n} \simeq 0.13$, the parity check is effectively a measurement of whether the resonator contains zero or one phonon. Each parity check, including qubit readout and conditional reset, completes in $\simeq12~\mu$s, more than two orders of magnitude faster than the phonon lifetime. We achieve a single-shot qubit readout fidelity of 0.91 (Fig.~\ref{fig:readout}). The single-shot parity assignment fidelity, $[P(e|\text{odd}) + 1-P(e|\text{even})]/2 = 0.68$, is limited primarily by qubit dephasing during the Ramsey sequence rather than by readout errors. In addition to parity checks, the strong dispersive regime enables direct spectroscopic measurement of the phonon number distribution: A spectrally narrow drive on the qubit resolves individual phonon numbers as peaks separated by $2\chi$ (Fig.~\ref{fig:spectra},~A and B), with the area under each peak yielding $P(n)$, the phonon number distribution~\cite{gambetta2006qubit,lee2023strong,von2022parity}.

\section*{Heralding mechanical Fock states}

\begin{figure}[tb]
\centering
\includegraphics[width=\columnwidth]{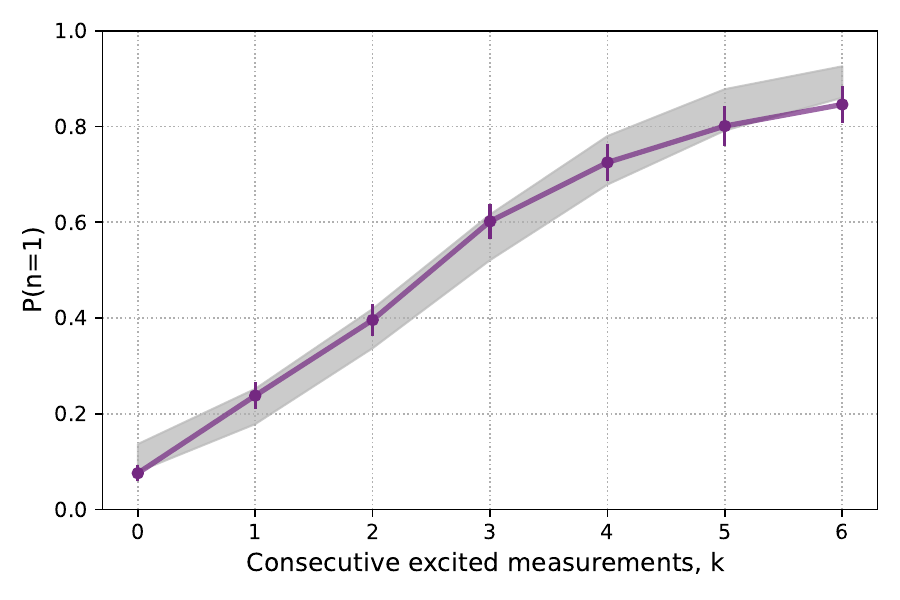}
\caption{\label{fig:Bayesian_update} \textbf{Evolution of $P(n\!=\!1)$ with postselection.} The purple circles illustrate $P(n\!=\!1)$ from the postselected qubit spectra in Fig.~\ref{fig:spectra}. Consecutive measurement outcomes of $\ket{e}$ increase $P(n\!=\!1)$ through projection of the mechanical state, reflecting the information we learn after each measurement outcome. The gray shaded region illustrates a fit obtained through forward-propagating $P(n)$ using Bayesian inference, where we fit $P(e|\text{even})$ and $\alpha$ of the initial coherent state. The error bars and shaded region represent 95\% confidence intervals of the data and fit respectively.}
\end{figure}

\begin{figure*}
\centering
\includegraphics[width=\textwidth]{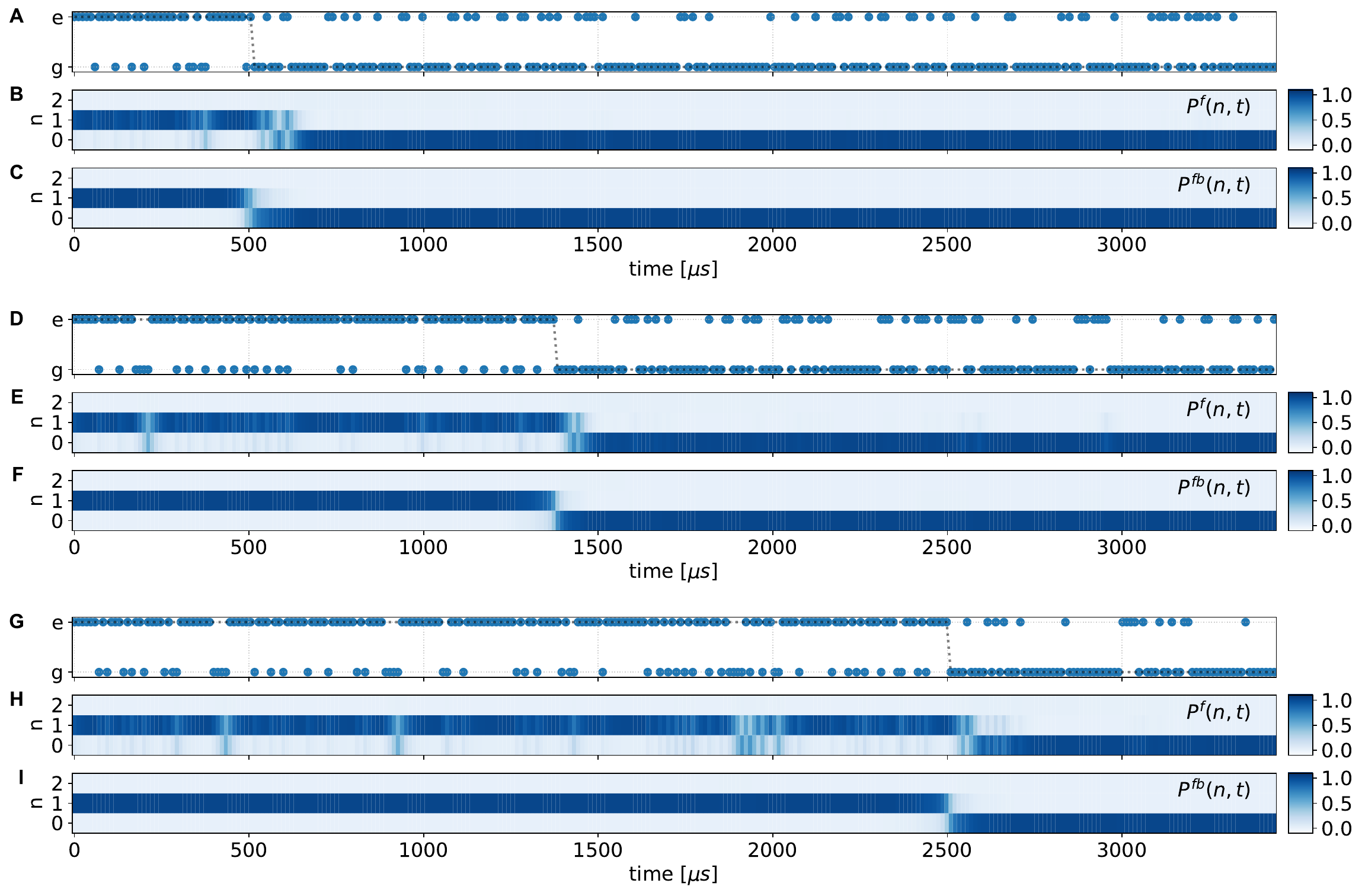}
\caption{\label{fig:trajectory} \textbf{Quantum jumps of a mechanical resonator.} Three individual measurement trajectories (rows), each consisting of 294 consecutive QND parity measurements on a mechanical resonator heralded in a single-phonon state with $P(n=1)=0.85$. \textbf{(A, D, and G)} Raw qubit measurement outcomes $q \in \{g, e\}$ after each Ramsey sequence. The dotted lines show the most-probable phonon number at each time step from forward-backward estimation. In each trajectory, an abrupt transition from predominantly $\ket{e}$ to predominantly $\ket{g}$ marks a quantum jump from $n=1$ to $n=0$. \textbf{(B, E, and H)} Forward-propagated phonon number distribution $P^f(n,t)$, using only measurements up to time $t$. \textbf{(C, F, and I)} Forward-backward distribution $P^{fb}(n,t)$, incorporating all measurements in the trajectory, which sharpens the jump and filters measurement noise. The jump occurs at a different time in each trajectory, which is consistent with the stochastic nature of quantum transitions.}
\end{figure*}

We demonstrate heralding of mechanical Fock states through repeated QND parity measurements. After preparing the mechanical mode in a weak coherent state with $\bar{n} \simeq 0.13$, we apply $k$ consecutive parity checks $\hat{\Pi}$ and then perform qubit spectroscopy to measure the phonon number distribution $P(n)$ (Fig.~\ref{fig:spectra}). Each consecutive measurement of $\ket{e}$ increases the weight in the $n=1$ peak, reflecting how each measurement sharpens our knowledge of the mechanical state (Fig.~\ref{fig:spectra}, C to H). After six consecutive measurements of $\ket{e}$, we herald a single-phonon state with a fidelity of $P(n=1) = 0.85$ (Fig.~\ref{fig:Bayesian_update}). The gradual increase of $P(n=1)$ with successive measurements reflects the modest single-shot parity fidelity of 0.68: Each measurement provides only partial information about the phonon number, so multiple rounds are needed to overcome measurement noise and build confidence in the state assignment.

This state evolution is quantitatively described by Bayesian forward propagation of the phonon number distribution~\cite{rybarczyk2015forward}. Starting from an initial distribution $P^f(n,0)$, each measurement outcome $q_{i+1} \in \{g,e\}$ updates the distribution according to~\cite{rabiner2002tutorial}
\begin{equation}\label{eq:Forward}
    P^f(n,t_{i+1}) = P(q_{i+1}|n) \, A \, P^f(n,t_i)
\end{equation}
where $P(q_{i+1}|n)$ is the probability of outcome $q_{i+1}$ given $n$ phonons, and the matrix $A$ accounts for phonon energy relaxation between measurements. We fit the measured evolution of $P(n=1)$ using this model with two free parameters: $P(e|\text{even})$, the probability of measuring $\ket{e}$ given an even phonon number, and $\alpha$, the amplitude of the initial coherent state. The fit (gray region in Fig.~\ref{fig:Bayesian_update}) shows good agreement with the data, yielding $\alpha = 0.356 \pm 0.087$ and $P(e|\text{even}) = 0.296 \pm 0.037$. We independently verify the fitted value of $\alpha$ using the data shown in Fig.~\ref{fig:spectra}B. From the relative peak areas corresponding to $n=0$ and $n=1$, we extract $\alpha=0.283\pm0.032$. The dominant source of measurement infidelity is qubit decoherence during the Ramsey sequence.

\section*{Tracking phonon number in real time}
Having established that our parity measurements project the mechanical state as expected, we now use them to continuously monitor the phonon number. \mbox{Figure}~\ref{fig:trajectory} shows three individual measurement trajectories, each consisting of 294 consecutive parity checks performed after heralding the resonator in a single-phonon state ($P(n=1) = 0.85$) through six consecutive measurements of $\ket{e}$, as in Fig.~\ref{fig:spectra}H. The raw qubit outcomes (Fig.~\ref{fig:trajectory}, A, D, and G) already reveal quantum jumps by eye: In each trajectory, an abrupt transition from predominantly $\ket{e}$ to predominantly $\ket{g}$ outcomes marks a jump from $n=1$ to $n=0$. A classical oscillator undergoing energy decay would instead produce a smooth, gradual shift in measurement statistics. To quantitatively extract $P(n,t)$ from these binary outcomes, we apply Bayesian state estimation to the full measurement record.

The forward distribution $P^f(n,t_i)$, computed from Eq.~\ref{eq:Forward} using measurements up to time $t_i$, provides an estimate of the phonon number using only past measurements (Fig.~\ref{fig:trajectory}, B, E, and H). Although it clearly resolves the quantum jumps, the forward estimate is sensitive to individual measurement fluctuations because each binary outcome directly updates the distribution. The forward-backward algorithm produces a smoother estimate by also incorporating measurements from times $t > t_i$. It combines the forward distribution with a backward likelihood $P^b(n,t_i)$, which propagates information from future measurements back to time $t_i$ (Appendix~\ref{app:forward_backward}), yielding~\cite{rabiner2002tutorial,rybarczyk2015forward}
\begin{equation}
    P^{fb}(n,t_i) = \frac{P^f(n,t_i)P^b(n,t_i)}{\sum_m P^f(m,t_i)P^b(m,t_i)} \; 
\end{equation}
The resulting probability distributions (Fig.~\ref{fig:trajectory}, C, F, and~I) show unambiguous quantum jumps between $n=1$ and $n=0$, occurring at a different time in each trajectory. This is analogous to photon jump trajectories observed in electromagnetic cavities~\cite{gleyzes2007quantum,sun2014tracking,rybarczyk2015forward} and single-photon detection through Bayesian estimation~\cite{dixit2021searching}, but now for a mechanical object.

During trajectories of repeated parity checks, residual measurement back action reduces the effective phonon lifetime to $T_\text{1eff} = 649~\mu s$ (Fig.~\ref{fig:mech_T1}). Our measurements are about 99\% quantum nondemolition: Comparing the intrinsic lifetime $T_1 = 2.1$ ms with the effective lifetime $T_{1\text{eff}} = 649~\mu$s during continuous monitoring yields a demolition probability of $P_D = 0.013$ per measurement (Appendix~\ref{app:mech_T1}). This is comparable to early photon jump experiments~\cite{gleyzes2007quantum} and could be reduced by better engineering the placement of the different transition frequencies.

By applying the forward-backward algorithm to an ensemble of such trajectories, we extract the distribution of times at which the resonator jumps from $n=1$ to $n=0$. This distribution is exponential with a time constant of $T_1 = 645 \pm 29~\mu$s (Fig.~\ref{fig:jump_statistics}A), which is consistent with the memoryless decay expected for spontaneous phonon emission. The extracted lifetime is robust to the choice of the input $T_1$ given to the forward-backward algorithm, varying by only $\pm 21~\mu$s as this parameter is swept from $570$ to $730~\mu$s (Fig.~\ref{fig:jump_statistics}B).

\section*{Conclusions}
The capability demonstrated here, fast repeated QND measurements of phonon number, is a consequence of our large cooperativity. The cooperativity, defined as $C = 4g^2 T_1T_2^*$, where $g$ is the coupling rate, $T_1$ is the mechanics lifetime, and $T_2^*$ is the qubit coherence time, roughly sets the number of coherent operations and measurements that can be performed within a single phonon lifetime. Our cooperativity exceeds $3\times10^6$, representing an order-of-magnitude improvement over other quantum acoustic platforms. The combination of strong, coherent coupling and long lifetimes is precisely the resource that underlies all bosonic quantum error correction. Nanomechanical resonators are compelling candidates for this application: Their compact size enables dense integration on-chip, their long lifetimes reduce the rate at which errors accumulate, and phonons' inability to propagate through vacuum provides natural isolation between modes~\cite{pechal2018superconducting,chamberland2022building}.

A concrete next step is a mechanical dual-rail qubit~\cite{teoh2023dual,de2025mid,koottandavida2024erasure}, in which two mechanical resonators are coupled to a shared qubit and joint parity measurements are used to detect errors. With this architecture, phonon loss would cause a detectable erasure error, while the effect of slow-frequency fluctuations can be canceled to first order by spatial overlap of mechanical modes to cause correlated dephasing or by effective decoupling through parametrically swapping the modes faster than the dephasing timescales~\cite{levine2024demonstrating}. Our transfer print integration technique enables this architecture because additional mechanical resonators can be stamped onto the same qubit.

Beyond quantum error correction, bringing real-time tracking of mechanical degrees of freedom into the quantum domain is important because vibrations couple to physical observables that are inaccessible to electromagnetic systems, such as mass, force, acceleration, and strain. For example, the combination of subpicogram effective masses and large quality factors is precisely what single-molecule mass sensing requires~\cite{naik2009towards,hanay2012single,hanay2015inertial,sage2018single}. The quantum control demonstrated here also adds a qualitatively new dimension to such sensing. The ability to prepare and nondestructively read out nonclassical mechanical states offers a route to sensing frequency changes at much lower drive power.

More broadly, a central theme in physics has been the extension of complete quantum control, progressing from state initialization through coherent manipulation and finally to QND measurements of quanta, to each new physical modality. Our work completes this arc for mechanical vibrations. These results bring the observation of quantum jumps, first seen in trapped ions~\cite{bergquist1986observation,nagourney1986shelved} and later in electromagnetic cavities~\cite{gleyzes2007quantum,sun2014tracking}, to a mechanical system, where the massive, vibrating objects that undergo these transitions offer a clear manifestation of the quantum nature of sound.

\begin{acknowledgments}
We acknowledge A. Hwang, S. Gyger, and K. Multani for helpful discussion. We thank IARPA and Lincoln Labs for providing us with the TWPA. We thank Amazon Web Services for their financial support. This material is based upon work supported by the Air Force Office of Scientific Research and the Office of Naval Research under award FA9550-23-1-0338. Any opinions, findings, and conclusions or recommendations expressed in this material are those of the author(s) and do not necessarily reflect the views of the United States Air Force or the Office of Naval Research. Device fabrication was performed at the Stanford Nano Shared Facilities (SNSF) and the Stanford Nanofabrication Facility (SNF), supported by NSF award ECCS-2026822. M.P.M. acknowledges support from a Natural Sciences and Engineering Research Council of Canada (NSERC) PGS-D fellowship. K.P. acknowledges support from the Department of Defense (DoD) through the National Defense Science \& Engineering Graduate (NDSEG) Fellowship Program.

\smallskip\noindent\textbf{Author contributions:} E.S. and T.M. designed, fabricated, and measured the device. E.S., T.M., O.A.H., M.P.M., K.P., and R.G.G.-F. developed the fabrication process for the mechanics device. E.S., T.M., M.P., and S.P.H. developed the fabrication process for the qubit device. E.S. and T.M. developed the aligned transfer print process. A.H.S.-N. and D.I.S. provided experimental and theoretical support. E.S., T.M., and A.H.S.-N. wrote the manuscript, with all others assisting. A.H.S.-N. supervised all efforts.

\smallskip\noindent\textbf{Competing interests:} A.H.S.-N. and D.I.S. are Amazon Scholars. The other authors declare no competing interests.

\smallskip\noindent\textbf{Data, code, and materials availability:} All data and software are available in the manuscript or the appendices or are deposited at Zenodo 10.5281/zenodo.20944616~\cite{zenodo}. Details of materials preparation can be found in Appendix~\ref{app:fabrication}.
\end{acknowledgments}

\appendix

\renewcommand{\thefigure}{S\arabic{figure}}
\renewcommand{\thetable}{S\arabic{table}}
\setcounter{figure}{0}
\setcounter{table}{0}

\section{Fabrication}
\label{app:fabrication}

We fabricate our mechanical resonator from X-cut 400 nm thick LN that is bonded to a 3 $\mu\text{m}$ layer of silicon oxide, all on a 500 $\mu\text{m}$ thick silicon carrier. We start by thinning our LN to 300 nm using a physical etch in an Argon ion mill. We then fabricate 75 nm spacers from LN that enable the one-dimensional phononic crystal to be suspended above the qubit. Specifically, spacer regions are protected using a hydrogen silsesquioxane (HSQ) mask patterned via electron-beam lithography and then etched in another Argon ion mill, leaving spacers of LN that are 75 nm taller than the remaining thin film. The HSQ mask is then stripped in a combination of hydrofluoric acid (HF) and piranha (2:1 $\text{H}_2\text{SO}_4$:$\text{H}_2\text{O}_2$). We then thin all LN regions by 75 nm using aggressive corrosive cleaning~\cite{gruenke2025surface}. Next, we pattern the phononic crystal cavity using electron-beam lithography with an HSQ mask and another Argon ion mill, resulting in a 150 nm thick mechanical resonator and 225 nm thick spacers. Finally, we release our resonators and the majority of the spacer area in dilute HF, which facilitates the aligned transfer print integration.

Our mechanical mode has a single-phonon $T_1$ that far exceeds similar previously reported devices~\cite{arrangoiz2019resolving,wollack2022quantum,lee2023strong}. We attribute this improvement to two key advances. First, there is no metal directly contacting the mechanical resonator, which has previously been reported to improve quality factors~\cite{wollack2021loss}. Second, the top surface of the resonator is never exposed to ion milling. Ion milling is known to lead to ion implantation and the creation of amorphous LN~\cite{gruenke2025surface}. We avoid ion milling the top surface of the LN and heal any damage caused by our ion mill etch to the sidewalls of the resonator via the final cleaning step of the  fabrication~\cite{gruenke2025surface}.

We fabricate our transmon from tantalum on silicon. We pattern the base layer, including ground planes and qubit islands, using electron-beam lithography with PMMA A8 as our resist. We do not use photolithography because we require a gap size of 500 nm in order to efficiently couple to the mechanical resonator. We etch our tantalum in a reactive ion etcher (RIE) with CF$_4$ chemistry and strip the remaining resist in solvents. We then acid clean our chip in a combination of piranha (2:1 $\text{H}_2\text{SO}_4$:$\text{H}_2\text{O}_2$) and dilute hydrofluoric acid (100 mL of $\text{H}_2\text{O}$ and 10 mL of 50\% HF). The Josephson junctions are fabricated using a Manhattan-style resist stack that enables junction deposition and bandaging without breaking vacuum and without ion milling the silicon substrate~\cite{tuokkola2025methods}. Liftoff is performed in N-methylpyrrolidone (NMP) which is heated to 80~$^\circ$C.

We integrate the mechanical resonator and qubit using an aligned transfer print technique. We use a FINEPLACER lambda flipchip die bonder to pick up the mechanics chip, align the mechanical resonators with the coupling regions of the qubit, and bond them together. The chips are bonded with a force of 10 N and are baked at 80~$^\circ$C for 5 minutes while the force is applied to ensure good adhesion. After 5 minutes, the bonder picks up the top chip, transferring the LN mechanical resonators directly to the qubit.

\begin{figure*}
\centering
\includegraphics[width=0.97\textwidth]{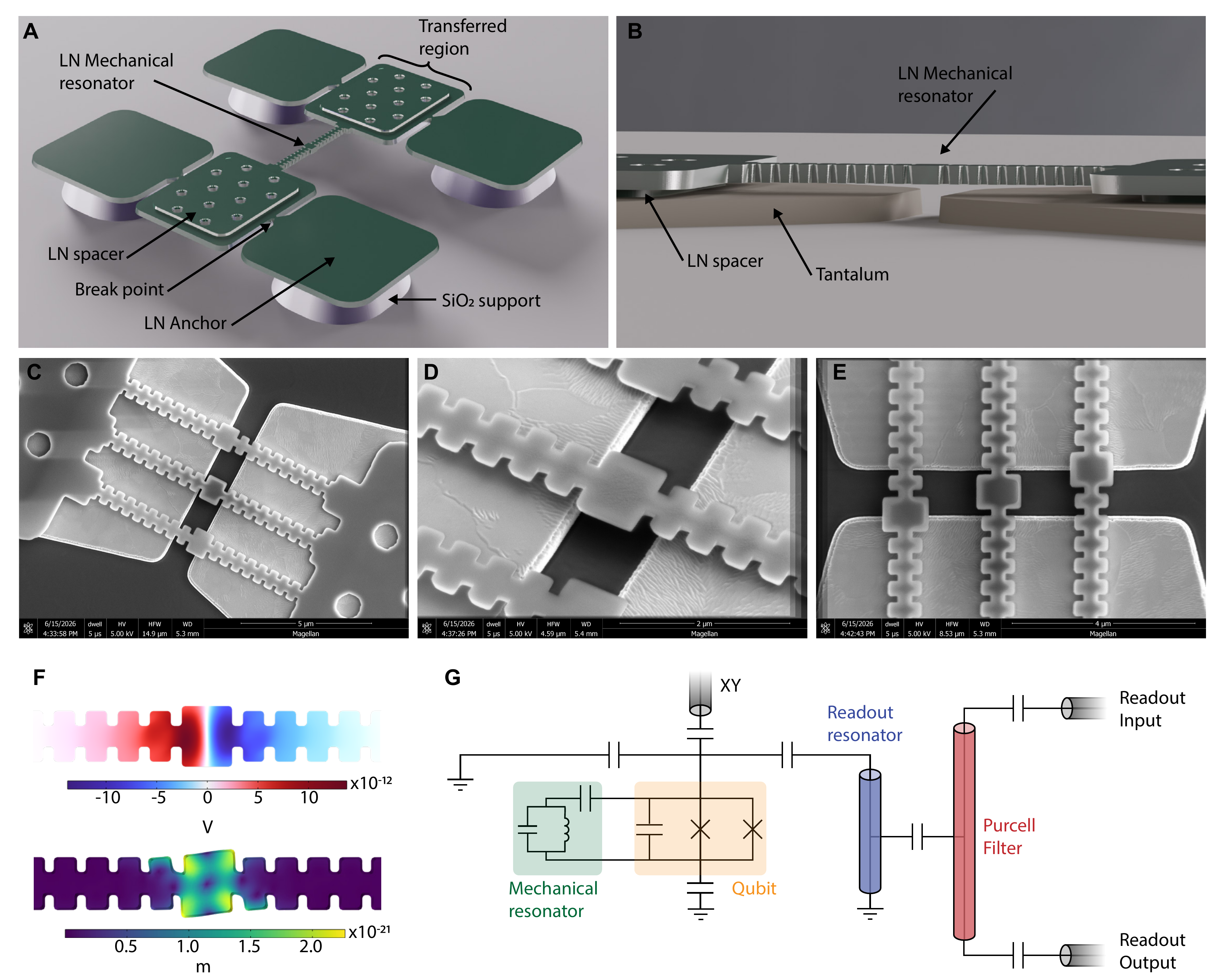}
\caption{\label{fig:device_details} \textbf{Device details.} \textbf{(A)} Illustration of the mechanical resonator chip immediately prior to the aligned transfer print. \textbf{(B)} Illustration of the mechanical resonator immediately after the aligned transfer print onto the qubit. \textbf{(C), (D), (E)} Scanning electron micrographs of the fully assembled device. \textbf{(F)} The simulated electrostatic potential (top) and displacement (bottom) for our mechanical resonator. \textbf{(G)} A simplified circuit diagram of our final device.}
\end{figure*}

In Fig.~\ref{fig:device_details}A we present an illustration of the mechanical resonator chip immediately prior to the aligned transfer print. The previous step involved etching the silicon dioxide layer in HF to release most of the LN area. Importantly, the resonator is completely released, as well as the raised spacers. The raised spacers have many holes etched through (via the earlier ion mill) to enable the HF to release the spacers. The four large-area pads labeled ``LN Anchor" are not fully released and are not intended to be transfer-printed onto the qubit. The spacers are attached to the anchors by thin LN bridges labeled ``Break point" that are intended to break during the transfer print process. In Fig.~\ref{fig:device_details}B we present an illustration of the mechanical resonator after it has been transfer printed onto the qubit. In Figs.~\ref{fig:device_details}C, D, E we show scanning electron micrographs of the device after aligned transfer printing.

We do not suspect that the aligned transfer print process negatively impacted the performance of the qubit. To control for potential losses introduced by integration, we measured a nominally identical qubit that did not see any transfer printing. This qubit was fabricated on the same chip as the one measured in this work. This control qubit had a $T_1$ of $61~\mu$s and a $T_2^*$ of $8.7~\mu$s. Compared to three other mechanics-integrated qubits on the same chip ($T_1 = 31, 81, 60\:\mu s$, $T_2^*=3.2, 10.5, 13.4\:\mu s$), the control qubit exhibits similar lifetimes and coherences. This implies that the limitations in qubit coherence originate from the qubit fabrication itself, and not from the aligned transfer print mechanics integration.

\section{Experimental wiring}
\label{app:wiring}
\begin{figure}
\centering
\includegraphics[width=\columnwidth]{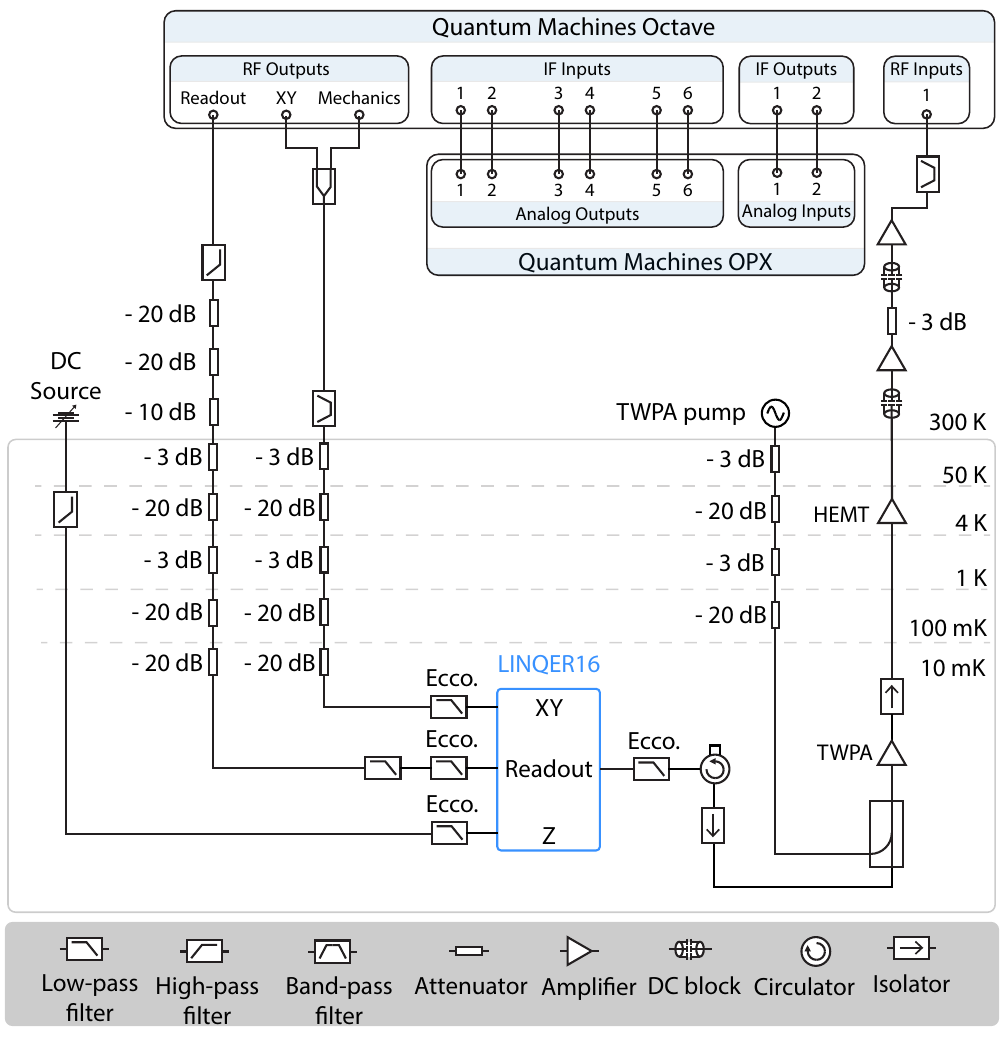}
\caption{\label{fig:wiring} \textbf{Experimental wiring diagram.}}
\end{figure}

Our experimental wiring is depicted in Fig.~\ref{fig:wiring}. All microwave control is achieved using an Operator-X (OPX) and Octave from Quantum Machines Inc. Microwave pulses (with frequencies below 200 MHz) are generated by the OPX Analog Outputs, which are then sent to the Octave IF Inputs. Within the Octave, these signals are up-converted and sent through the Octave RF Outputs. The I and Q quadratures used to drive the readout resonator are generated by the OPX Analog Outputs 1 and 2. The I and Q quadratures used for qubit gate pulses are generated by the OPX Analog Outputs 3 and 4. Finally, the I and Q quadratures used to drive the mechanical resonator are generated by the OPX Analog Outputs 5 and 6. Similarly, microwave signals emitted from our device are sent to the Octave RF Inputs, down-converted, then transferred from the Octave IF Outputs to the OPX Analog Inputs. These signals are then digitized by the OPX, where they are then processed and saved into memory.

To achieve fast, single-shot measurements, we read out our qubit through a traveling-wave parametric amplifier (TWPA) from Lincoln Labs~\cite{macklin2015near}. To isolate our qubit from TWPA noise, we use a combination of an isolator and a terminated circulator. We are careful to use a cryogenic 50 $\Omega$ termination that is well-thermalized to the base plate of our dilution refrigerator.

We use a DC voltage source (SRS SIM928) to bias our tunable qubit. The flux line is low-pass filtered at the 4 K stage (Aivon Therma-24G).

Our qubit is packaged in a LINQER16 from \mbox{SCALINQ}. Before the LINQER16, all lines are filtered using Eccosorb infrared radiation filters.

\section{Device details and characterization}
\label{app:parameters}

\begin{figure}
\centering
\includegraphics[width=\columnwidth]{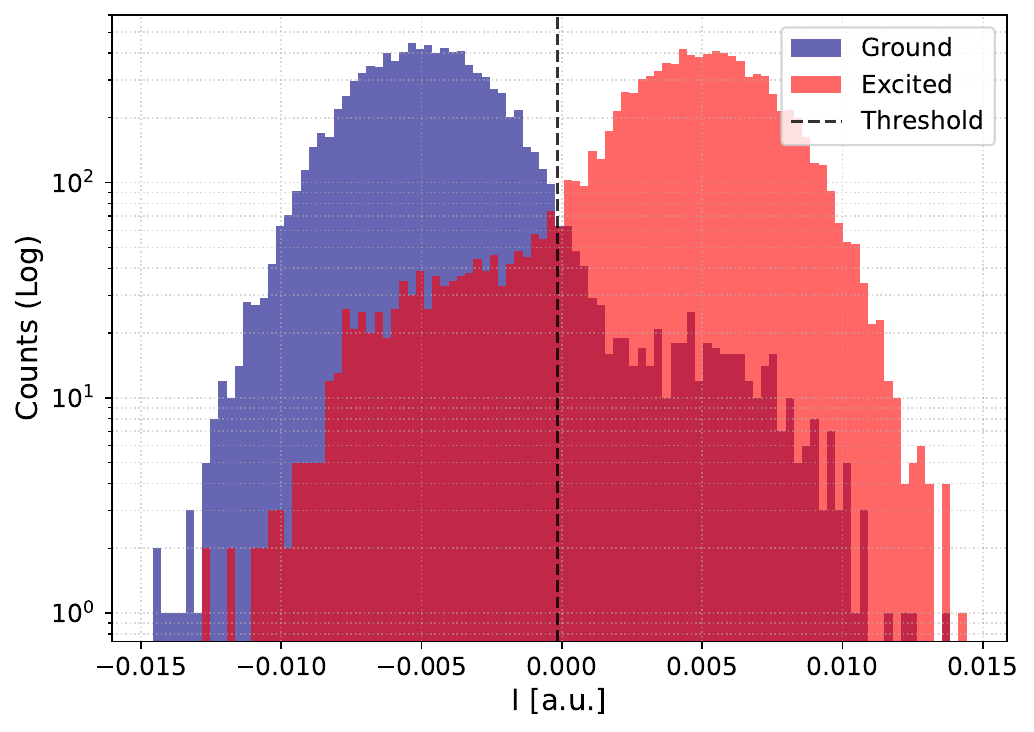}
\caption{\label{fig:readout} \textbf{Single-shot readout fidelity.} We achieve a single-shot readout fidelity of $0.912$ using a 1792 ns square pulse.}
\end{figure}

Our device properties are summarized in Table~\ref{table:sup_device_params}. Our single-shot readout performance is characterized in Fig.~\ref{fig:readout}. We achieve an average readout fidelity of $F =$ 0.912. Our readout pulse is a square pulse with a total duration of 1792 ns.

We flux bias our qubit such that our qubit and mechanics are in the strong dispersive regime. Specifically, we bias our qubit such that the mechanical mode is in the straddling regime of the qubit spectrum, with the mechanical mode detuned by roughly 21 MHz from the qubit ef transition. We characterize the shift-per-phonon in the qubit spectrum $2\chi$ by displacing the mechanical mode and then performing qubit spectroscopy, as in Fig.~\ref{fig:spectra}. We observe a number splitting of $2\chi/2\pi=$ 328 kHz.

\begin{figure*}
\centering
\includegraphics[width=0.7\textwidth]{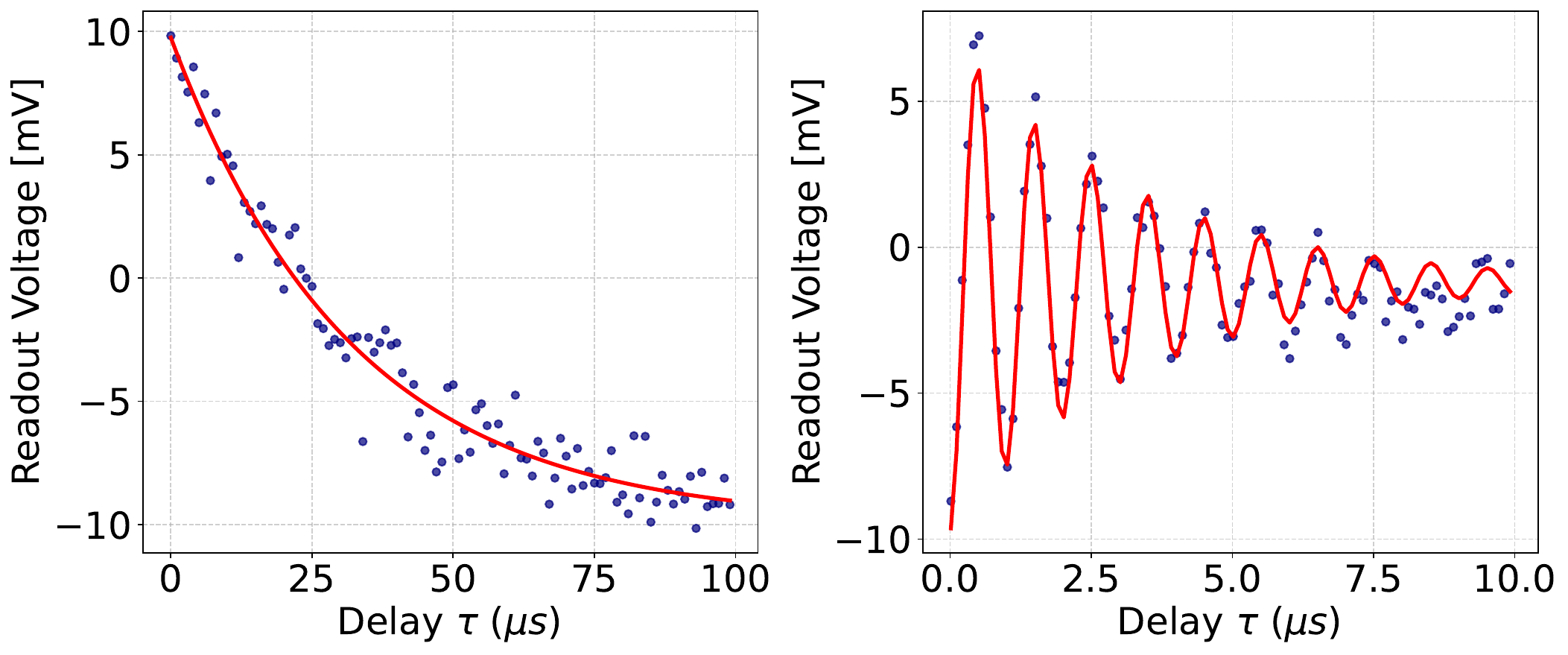}
\caption{\label{fig:qubit_T1_T2} \textbf{Qubit characterization.} Qubit lifetime (left) and coherence (right). }
\end{figure*}

Our qubit coherences at this flux bias are shown in Fig.~\ref{fig:qubit_T1_T2}. We measure the lifetime of our qubit $T_{1q}$ by driving the qubit with two sequential $\pi/2$ pulses, followed by measuring the qubit state after a variable delay. Our gate $\pi/2$ pulses are Gaussian pulses with a temporal standard deviation of 11.2 ns. We measure the Ramsey coherence time of our qubit $T_2^*$ by driving our qubit with a $\pi/2$ pulse, waiting a variable delay, driving our qubit with a $-\pi/2$ pulse, then finally measuring the qubit state.

Our mechanical resonator mode is a shear mode, similar to previous experiments~\cite{arrangoiz2019resolving,wollack2022quantum,lee2023strong}. In Fig.~\ref{fig:device_details}F we show the simulated electrostatic potential (top) and displacement (bottom) for our mode. The effective mass of our mode is $m_{\text{eff}}=0.1995 $ pg, yielding a zero-point fluctuation magnitude of $x_{\text{zpf}}=3.576$ fm. The electrostatic potential profile shown in Fig.~\ref{fig:device_details}F helps illustrate how we couple our mechanical resonator to our qubit. By placing our qubit electrodes near the resonator via aligned transfer print, the oscillating voltage from the piezoelectric effect drives voltage oscillations on our qubit islands.

In Fig.~\ref{fig:device_details}G we show a simplified circuit diagram. Our qubit is a tunable transmon qubit (highlighted in orange) and the mechanical resonator can be modeled by an effective LC circuit (highlighted in green)~\cite{wollack2021loss, lee2023strong}. The qubit is capacitively coupled to a readout resonator, which is coupled to a Purcell filter. The coupling between the readout resonator and Purcell filter is both capacitive and inductive. We extract the full coupling via finite-element electromagnetic simulations.

\begin{figure}
\centering
\includegraphics[width=\columnwidth]{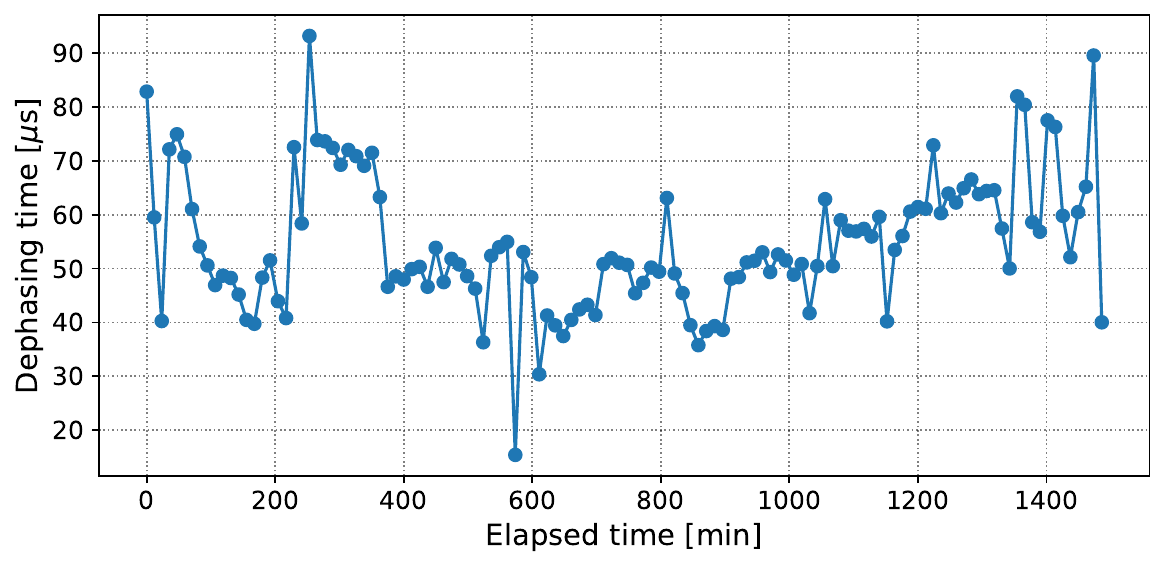}
\caption{\label{fig:mech_T2} \textbf{Mechanics dephasing time.} We measured the mechanical dephasing time of a nominally identical mechanical resonator, following the protocol described in reference~\cite{cleland2024studying}. All measurements were conducted at single-phonon powers. We tracked the mechanical dephasing time over a period of roughly 24 hours and observe large fluctuations. The average dephasing time is $55~\mu$s.}
\end{figure}

We did not measure the dephasing time for the mechanical resonator reported in this paper. However, we did measure the dephasing time for a nominally identical mechanical resonator that was fabricated on the same chip and transfer printed onto a nominally identical qubit. We measured our mechanical dephasing time following the protocol described in reference~\cite{cleland2024studying} and observed that it varies significantly over time. In Fig.~\ref{fig:mech_T2}, we plot the mechanical dephasing time, measured at single-phonon powers, over a period of roughly 24 hours. The frequency noise and dephasing properties of quantum acoustic systems are an active area of research~\cite{cleland2024studying,bozkurt2025mechanical}, and two-level system (TLS) defects are widely considered to play a dominant role in dephasing. In the present work, we did not carefully investigate the TLS properties of the mechanical resonator.

\begin{table}
\centering
\begin{tabular}{||c| c ||}
 \hline
 $\omega_\text{ge}/2\pi$& 3.389 GHz\\
 \hline
 $T_{1q}$ & 31 $\mu\text{s}$ \\
\hline
 $T_2^*$ & 3.2 $\mu\text{s}$ \\
 \hline
 $\alpha/2\pi$ & 133 MHz \\
 \hline
 $\omega_\text{m}/2\pi$& 3.277 GHz\\
 \hline
 $T_1$ & 2.1 ms \\
 \hline
 $\chi/2\pi$ & 164 kHz \\
  \hline
 $m_\text{eff}$ & 0.1995 pg \\
  \hline
 $x_\text{zpf}$ & 3.576 fm \\
  \hline
\end{tabular}
\caption{Table of device parameters. The qubit parameters $\omega_\text{ge}$, $T_{1q}$, $T_2^*$, and $\alpha$, as well as the dispersive shift $\chi$ are all reported at the flux bias where the data from Fig.~\ref{fig:spectra} and Fig.~\ref{fig:trajectory} were taken. The measurement of the mechanical lifetime $T_1$ is described in Appendix~\ref{app:mech_T1}. The effective mass $m_\text{eff}$ and zero-point fluctuations $x_\text{zpf}$ are also reported.}
\label{table:sup_device_params}
\end{table}

\section{Mechanics lifetime and measurement backaction}
\label{app:mech_T1}
\begin{figure}
\centering
\includegraphics[width=\columnwidth]{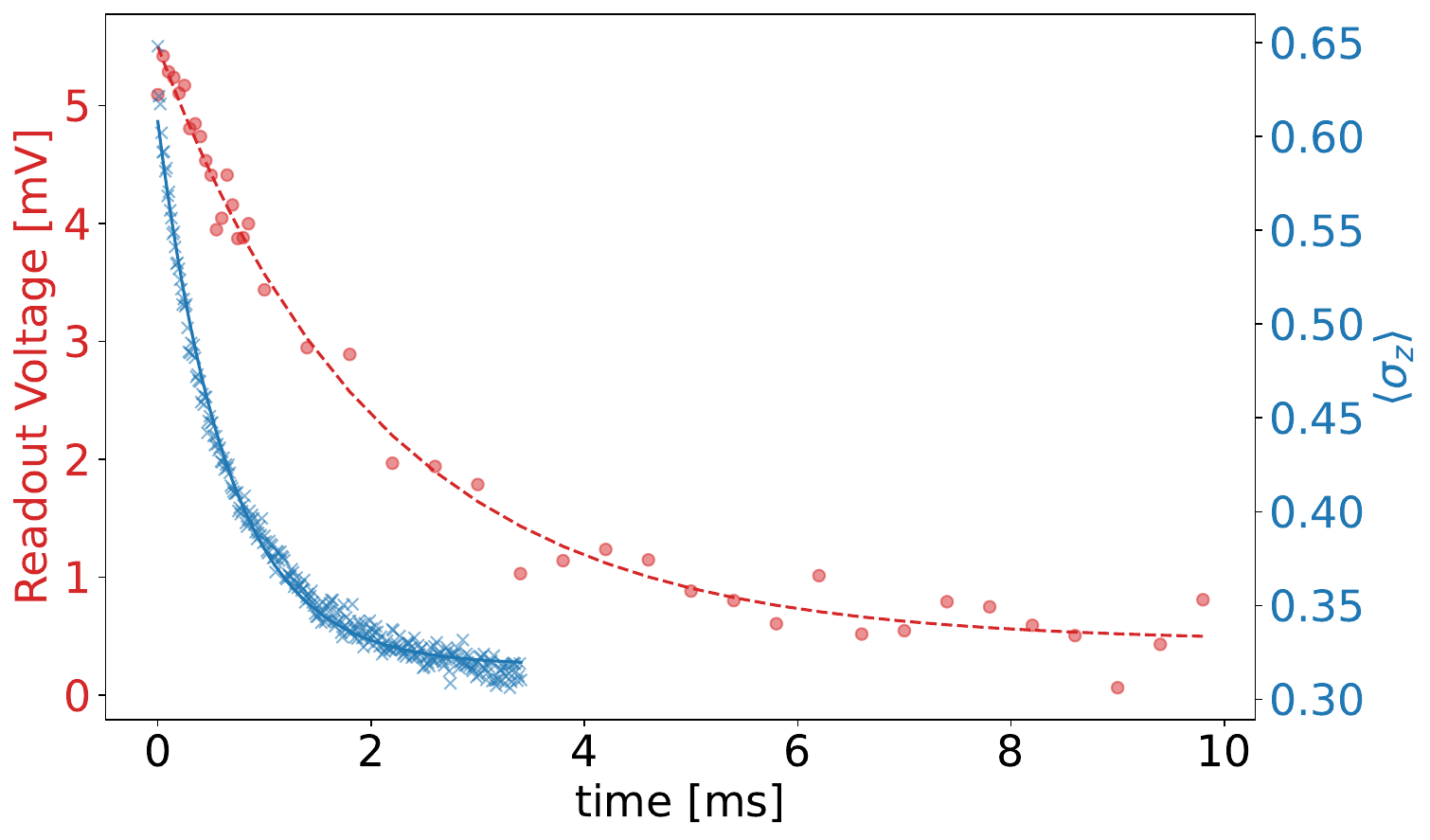}
\caption{\label{fig:mech_T1} \textbf{Mechanics lifetime.} Red circles: Measuring the lifetime of an $\bar{n}=1$ coherent state via qubit spectroscopy. We vary the time between the mechanics excitation pulse and the qubit spectroscopy pulse. A red dashed line shows an exponential fit, yielding $T_1 = 2.1$ ms. Blue crosses: qubit state $\braket{q}$ averaged over many phonon number-tracking trajectories, postselected on $k=6$ consecutive $\ket{e}$ measurements. A blue solid line shows an exponential fit, yielding an effective lifetime of $T_\text{1eff}=649~\mu\text{s}$.}
\end{figure}

We characterize the intrinsic mechanical lifetime by preparing a coherent state with $\bar{n}=1$ and monitoring the $n=0$ qubit spectral line after a variable delay (red circles in Fig.~\ref{fig:mech_T1}). An exponential fit yields $T_1 = 2070 \pm 240~\mu$s.

During repeated parity measurements, the mechanical mode decays significantly faster due to residual measurement backaction. To quantify this, we average the qubit measurement outcome $\braket{q}$ over many trajectories that are heralded in a single-phonon state via six consecutive $\ket{e}$ measurements as in Fig.~\ref{fig:spectra}H. This ensemble averaged decay of $\langle q \rangle$ is shown by blue crosses in Fig.~\ref{fig:mech_T1}. The resulting decay gives an effective lifetime of $T_\text{1eff} = 649~\mu$s, roughly three times shorter than the intrinsic value. Following~\cite{sun2014tracking}, we model this as a per-measurement excess decay probability $P_D$:
\begin{equation}
    1/T_{1\text{eff}} = 1/T_1 + P_D/\tau_{\text{rep}}
\end{equation}
where $\tau_{\text{rep}} \simeq 12~\mu$s is the measurement repetition period. This yields $P_D \simeq 0.013$, indicating that each parity check preserves the phonon number with ${\sim}99\%$ probability.

One possible explanation for our measurement backaction is unintended swapping between the $|e,1\rangle$ and $|f,0\rangle$ states. Given that our mechanical mode is 21 MHz detuned from the qubit $e\rightarrow f$ transition and that our mechanical mode and qubit are coupled at a rate of $g/2\pi=1.72$ MHz, we can simulate the time-averaged swap probability between the mechanical mode and the qubit $e\rightarrow f$ transition during the Ramsey protocol. A numerical simulation of the master equation yields a swap probability of roughly 1.6\%. Furthermore, between parity measurements we apply conditional $f$-state reset with a fidelity of roughly 67.1\%, indicating that the probability that we demolish a phonon via swapping to the $f$-state followed by qubit reset is roughly 1.1\%. This value is consistent with the observed demolition probability $P_D\simeq 0.013$ reported above. In the future, this demolition probability can be further reduced with more careful engineering of the qubit and mechanical resonator frequencies.

\begin{figure}[t]
\centering
\includegraphics[width=0.97\columnwidth]{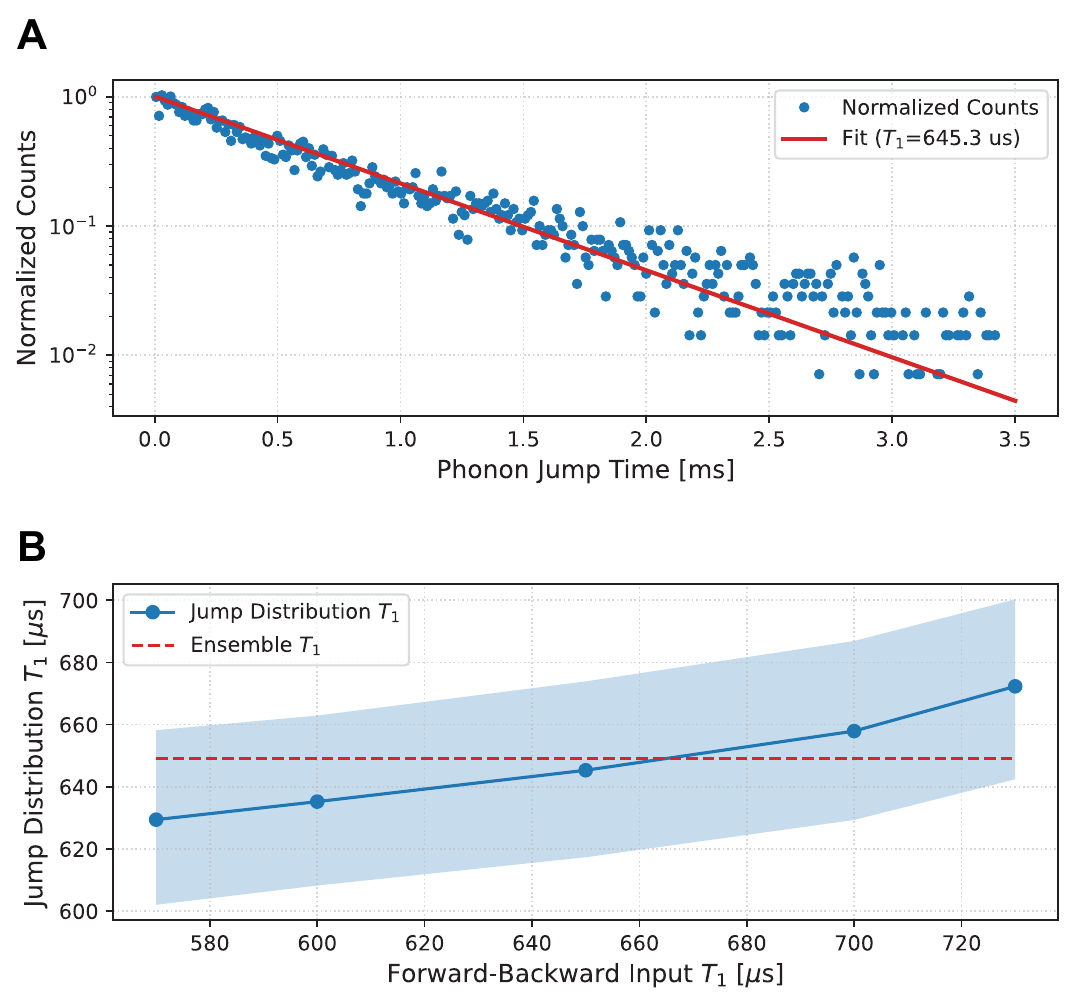}
\caption{\label{fig:jump_statistics} \textbf{Phonon jump time distribution.} \textbf{(A)} The blue circles show the distribution of when the mechanical resonator jumps from $n=1$ to $n=0$ according to $P^{fb}(n,t)$. The red line is a single-exponential fit with a $T_1$ of $645~\mu s\pm 29~\mu s$. \textbf{(B)} We confirm that our forward-backward algorithm does not simply reproduce the input $T_1$ by sweeping this value from $570~\mu s$ to $730~\mu s$ and recalculating the jump distribution. The predicted $T_1$ from the jump distribution is shown by blue circles, highlighting the relative insensitivity to the forward-backward algorithm input. The effective lifetime $T_\text{1eff}$ from Fig.~\ref{fig:mech_T1} is shown by a red dashed line.}
\end{figure}

\section{Forward-backward state estimation}
\label{app:forward_backward}
The forward distribution $P^f(n,t_i)$ is computed by iterating Eq.~\ref{eq:Forward} forward in time from an initial prior $P^f(n,t_0)$. The backward likelihood $P^b(n,t_i)$ captures the information contained in all measurements after time $t_i$. It is initialized as a uniform distribution at the final time step and propagated backward according to~\cite{rabiner2002tutorial}
\begin{equation}
    P^b(n,t_i) = A^T P(q_{i+1}|n) \, P^b(n,t_{i+1})
\end{equation}
where $A^T$ is the transpose of the decay matrix and $P(q_{i+1}|n)$ is the emission probability defined in Eq.~\ref{eq:Forward}. The forward-backward estimate is then obtained by combining the two distributions as described in the main text.

We use the forward-backward algorithm to smooth data postselected on $k=6$ consecutive measurements of $\ket{e}$ as in Fig.~\ref{fig:spectra}H. This postselection condition allows us to smooth all subsequent points in the trajectory assuming an initial single-phonon state with $P^f(n, t_0) = (0.15, 0.85, 0, 0, \dots)$, obtained from number splitting measurements (see Fig.~\ref{fig:spectra}). The forward-backward algorithm incorporates the effect of our measurements, $P(e|\text{even})=0.296$ and $P(e|\text{odd})=0.653$, with values obtained from the fit in Fig.~\ref{fig:Bayesian_update}. Since the algorithm also takes into account intrinsic decay of the mechanical resonator, it also takes an input phonon lifetime $T_1$ and thermal occupation $n_{bath}$. We estimate $n_{bath}$ through the average excited state population of the qubit with no coherent drive on the mechanics, as $P(e)=P(e|\text{even})P(n=0) + P(e|\text{odd})P(n=1)$, which gives $n_{bath}=0.0128$. The input phonon lifetime is derived from the effective lifetime $T_\text{1eff}$ shown in Fig.~\ref{fig:mech_T1}. This thermal bath temperature corresponds to a temperature of 36 mK for our mechanical mode. From measuring the excited-state thermal population of the qubit without any drives, we measure our qubit temperature to be 53 mK. The difference in temperatures is likely due to different loss channels and baths associated with acoustic versus electromagnetic resonators.

Using $P^{fb}(n,t)$ we can assign a most likely phonon number at any given time in the trajectory. This allows us to determine the times at which the mechanical resonator jumps from $n=1$ to $n=0$, which is shown in Fig.~\ref{fig:jump_statistics}. The distribution of jump times is single-exponential with a $T_1=645~\pm~29~\mu$s, demonstrating good agreement with the $T_\text{1eff}=649~\mu$s. To verify that the jump distribution does not simply reproduce the input $T_1$ given to the forward-backward algorithm, we vary this value across a wide range $T_1 \in [570~\mu$s$, 730~\mu$s$]$. We find that the lifetime from the jump distribution only varies by $\pm21~\mu$s as we sweep the input $T_1$ that we use in the forward-backward algorithm, as shown in Fig.~\ref{fig:jump_statistics}.

Our parity interrogation cannot differentiate between $n=0$ and $n=2$ phonons in the resonator. Therefore, given that our forward prior is $(0.15, 0.85, 0, 0, ...)$, trajectories that start with repeated even parity measurements are categorized as starting in $n=0$ and not $n=2$. However, given that we are working with coherent states where $\bar{n}\simeq0.13$ we have $P(0) = 0.878$, $P(1) = 0.114$, and $P(2) = 0.00742$. Therefore, $n=2$ events are rare.

\clearpage 
\bibliography{phonon_jumps}

\end{document}